\documentclass[conference]{IEEEtran}
\usepackage{cite}
\usepackage{booktabs}
\usepackage{amsmath,amssymb,amsfonts}
\usepackage{algorithmic}
\usepackage{graphicx}
\usepackage{textcomp}
\usepackage{xcolor}
\makeatletter
\newcommand{\newlineauthors}{%
  \end{@IEEEauthorhalign}\hfill\mbox{}\par
  \mbox{}\hfill\begin{@IEEEauthorhalign}
}
\makeatother
\def\BibTeX{{\rm B\kern-.05em{\sc i\kern-.025em b}\kern-.08em
    T\kern-.1667em\lower.7ex\hbox{E}\kern-.125emX}}

\begin{document}
\title{Anxiety Detection Leveraging Mobile Passive Sensing}

\author{
\IEEEauthorblockN{Lionel M. Levine}
\IEEEauthorblockA{
\textit{UCLA}\\
Los Angeles, CA \\
lionel@cs.ucla.edu}
\and
\IEEEauthorblockN{Migyeong Gwak}
\IEEEauthorblockA{
\textit{UCLA}\\
Los Angeles, CA \\
mgwak@cs.ucla.edu}
\and
\IEEEauthorblockN{Kimmo K\"arkk\"ainen}
\IEEEauthorblockA{
\textit{UCLA}\\
Los Angeles, CA \\
kimmo@cs.ucla.edu}
\and
\IEEEauthorblockN{Shayan Fazeli}
\IEEEauthorblockA{
\textit{UCLA}\\
Los Angeles, CA \\
shayan@cs.ucla.edu}
\newlineauthors
\IEEEauthorblockN{Bita Zadeh}
\IEEEauthorblockA{
\textit{Chapman University}\\
Orange County, CA \\
bzadeh@chapman.edu}
\and
\IEEEauthorblockN{Tara Peris}
\IEEEauthorblockA{
\textit{UCLA}\\
Los Angeles, CA \\
tperis@mednet.ucla.edu}
\and
\IEEEauthorblockN{Alexander S. Young}
\IEEEauthorblockA{
\textit{UCLA}\\
Los Angeles, CA \\
ayoung@ucla.edu}
\and
\IEEEauthorblockN{Majid Sarrafzadeh}
\IEEEauthorblockA{
\textit{UCLA}\\
Los Angeles, CA \\
majid@cs.ucla.edu}
}

\maketitle

\begin{abstract}
Anxiety disorders are the most common class of psychiatric problems affecting both children and adults. However, tools to effectively monitor and manage anxiety are lacking, and comparatively limited research has been applied to addressing the unique challenges around anxiety. Leveraging passive and unobtrusive data collection from smartphones could be a viable alternative to classical methods, allowing for real-time mental health surveillance and disease management. This paper presents eWellness, an experimental mobile application designed to track a full-suite of sensor and user-log data off an individual's device in a continuous and passive manner. We report on an initial pilot study tracking ten people over the course of a month that showed a nearly 76\% success rate at predicting daily anxiety and depression levels based solely on the passively monitored features.
\end{abstract}


\section{Background and Introduction}

%Effective monitoring of mental health is a critical priority. It is estimated that nearly one in five U.S. adults lives with a mental illness \cite{theeke2009predictors}, and 11.2 million adults (or 4.5\% of all U.S. adults) experienced some form of serious mental illness that resulted in severe functional impairment\cite{dhodapkar2020survey}. 

Within the spectrum of mental health disorders, Anxiety disorders are the most common class of psychiatric problems affecting both children and adults \cite{bitsko2018epidemiology} \cite{cartwright2006anxiety}\cite{merikangas2010lifetime}, with up to one in three people in the US meeting full diagnostic criteria by early adulthood \cite{topper2017prevention}\cite{hammerness2008characterizing}. This breaks down roughly to 7 to 9\% suffering from a specific phobia, 7\% from social anxiety disorder, and 2 to 3\% each from panic disorder, agoraphobia, generalized anxiety disorder, and separation anxiety disorder.\cite{APA} Individuals with anxiety disorders contend with substantial distress and impairment. They are at heightened risk for a host of negative long-term outcomes including depression, substance use, educational underachievement, and poor physical health \cite{newby2014effectiveness}\cite{bardone1998adult}\cite{woodward2001life}

The optimal method for the prevention or care of mental illness is early identification, diagnosis, and proactive treatment\cite{wagner2016mental}. Time-sensitive intervention is crucial for preventing conditions from becoming chronic and debilitating. However, traditional methods of psychiatric assessment, including clinical interviews and self-reports, are limited in their ability to provide just-in-time intervention as well as early identification. They depend heavily on retrospective summaries collected in clinical settings, conditions that often result in reporting biases, inaccurate recall, or late and ineffectual treatment.

Additionally, anxiety disorders are, for the most part, vastly overlooked and under-treated in the community; only 15-30\% of anxious individuals in the community receive treatment of any kind. Recent research has found strikingly high levels of anxiety among college-age youth. Indeed, 58.4\% of college-aged youth report feeling “overwhelmed by anxiety” \cite{american2016american}.  Several other recent studies document the high proportion of college students meeting full diagnostic criteria for an anxiety disorder \cite{bruffaerts2018mental}. At the same time, young adults are particularly overlooked within the health care system, with rates of screening, identification, and referral falling below those of either children or adults \cite{woodward2001life}. Given this landscape, there remains a pressing need for tools that improve early identification of anxiety symptoms, provide users with the tools to monitor their activities, raise awareness of factors impacting on their wellbeing, and provide a mechanism for intervention should an anxiety episode escalate.

The growing ubiquity of consumer devices, among them smartphones, smartwatches, and in-home sensors, all equipped with an array of sensors and user-logs, have resulted in an unprecedented opportunity to catalog and quantify the daily aspects of an individual’s life, creating repositories of personalized information \cite{swan2012health}. 

While much has been noted about the insidious aspects of such surveillance capabilities, there is also significant potential for such monitoring, if harnessed and utilized by the individuals themselves, to dramatically improve their healthcare outcomes. Such tools could potentially allow the user to accurately track their behaviors and habits, compare personal activities with population-level baselines, establish outlier behaviors with their peers, and even motivate behavioral change and the promotion of healthy habits. 

There is significant potential for such monitoring if harnessed and utilized by the individuals themselves, to improve their healthcare outcomes dramatically. While physical behavior and physiological health are extensively tracked, mental health is largely overlooked. 
%The notable exception to this is physiological stress monitoring of features like heart-rate variability and Galvanic Skin Response that is accomplished by wearable sensing devices like smartwatches to determine stress level \cite{ciabattoni2017real}. 

Specifically, The capability to track behavioral metrics and associate them to mental health, although intimately linked, has not been definitively established. This owes to the significant difficulty in correlating monitorable behaviors and corresponding mental health. Behavioral patterns both within (e.g., the transition from weekday to weekend) and across individuals (e.g., simple differences in how many men and women carry their phones) are simply too diverse and too subject to confounding factors beyond mental health to allow for easy correlations. Nevertheless, the growing challenges around mental health, necessitate exploring the possibility further. 

Recent efforts have explored whether pervasive mental health monitoring could be feasible through a smartphone and the embedded sensors, such as motion sensors, ambient light, microphone, camera, Global Positioning System (GPS), proximity, and touch screen\cite{ciabattoni2017real}\cite{osma2014proposal}\cite{ben2015next}\cite{narziev2020stdd}. These efforts have shown the promise of this approach in successfully tying behavioral monitoring to mental health; however, such approaches have not translated into fully mature frameworks, and have focused almost exclusively on depression-related conditions, which while often spoken in conjunction with anxiety, manifest in distinct ways\cite{fukazawa2019predicting}.

The advantages of leveraging a smartphone-based platform are that the continuous and quantitative collection of data potentially provides a more reliable indicator of an individual's risk at any given time, as well as offering a mechanism for just in time intervention should a mental health episode occur\cite{ben2015next}. Conversely, smartphone-derived data present several challenges, some of which have already been noted, which can result in limited accuracy owing to differences in behavioral patterns across users, and the indirect manner of detection\cite{fukazawa2019predicting}.

We present a system for the remote monitoring of mental health symptoms, their fluctuation, and their attendant disruption to personal functioning, called eWellness. The eWellness framework is designed to capture a broad spectrum of remote monitoring, survey data acquisition, secure data transmission and management, data analytics, and visualization.

The primary component of eWellness is a mobile application that facilitates data collection and transmission harvested from an array of sensors and usage logs from a user’s smartphone. The data is collected passively, pre-processed, and transmitted through a secure gateway to the cloud, where it is securely stored, and indexed using a scalable database.

Concurrently the eWellness application includes an active querying component where users can be prompted with Ecological Momentary Assessments (EMA) of their mental health status. This architecture is complemented by a back-end analytic engine, capable of mapping observed metrics and exogenous data sources to a user’s mental health state, based on adaptive statistical models, and advanced machine learning algorithms. The system is designed to monitor overall mental health as well as acute crisis events in both a retrospective and predictive capacity.

\section{Framework}

\subsection{Server}
Data from the study, both usage-logs and EMAs response, are first encrypted and cached locally on the user's device, and then transmitted to a secure remote server, where it is stored in an encrypted scalable MySQL database. 

\subsection{eWellness Data Collection}
The eWellness mobile application, developed for android devices, collects passive behavioral data derived from user-logs and embedded sensors, capturing the following metrics:
\begin{itemize}
\item \textbf{Communication}: is monitored by incoming and outgoing phone calls and text messages, including the duration of phone calls, the number of texts and phone calls, and unique individuals contacted. This does not assess the content of communications or the recipient of the communication, beyond establishing a unique contact.
\item \textbf{Location}: is periodically sampled using GPS, network, and Wi-Fi detection. Prompts for a new location after moving 5 meters, up to once a minute. This metric leverages the Google Fused Location API. The application does not track specific locations; instead, it keeps a total distance traveled using the vectorized haversine distance function.
\item \textbf{Ambient Sound}: detects speech and communication above 50 decibels using the phone’s microphone. It monitors every 5 minutes for 5 seconds. This metric does not assess the content of communications and merely records the sound frequency and decibel level in a numeric value.
\item \textbf{Activity and Movements}: leverage the device’s accelerometer, gyroscope, and GPS tracking. Activity is sampled every 60 seconds. In order to determine stationary and moving activity-type, the application leverages Google’s Activity Recognition API.
\item  \textbf{Light}: detects light level associated with possibly being in an outdoor or indoor location. This sensor is sampled every 6 seconds. 
\item \textbf{Phone use}: is user-log monitoring the device's screen on-time. 
\end{itemize}

We derived daily aggregated features from these metrics to define the relationship between smartphone sensors and anxiety symptom severity. We obtained statistical characteristics, such as minimum, maximum, mean, standard deviation, the 25th, 50th, and 75th percentiles, of the numeric values of noise exposure and the ambient luminance. The number of activity transitions and duration of each physical activity per day also became a significant metric of identifying mentally distressed days. 

\subsection{Limiting Personally Identifiable Data Collection}
Recognizing the potentially invasive nature of applications like this, data collection was carefully scoped to avoid the collection of Personally Identifiable Information (PII) that could link a particular user to a particular dataset. For example, when attempting to gauge sociability, the application logs the total number of phone calls made, total time on the phone, and the number of unique contacts called; the identities of specific callers were not tracked. This has the consequence of introducing a degree of obscurity into an observed finding (e.g., as the application is unable to differentiate between calls to friends and calls to a customer-service hotline). At the same time, in the interest of both respecting privacy and ensuring the acceptability of the app, these efforts were felt to be necessary constraints on data collection.

\section{Pilot Study Methodology}

\begin{figure}[h]
  \centering
  \includegraphics[width=0.8\linewidth]{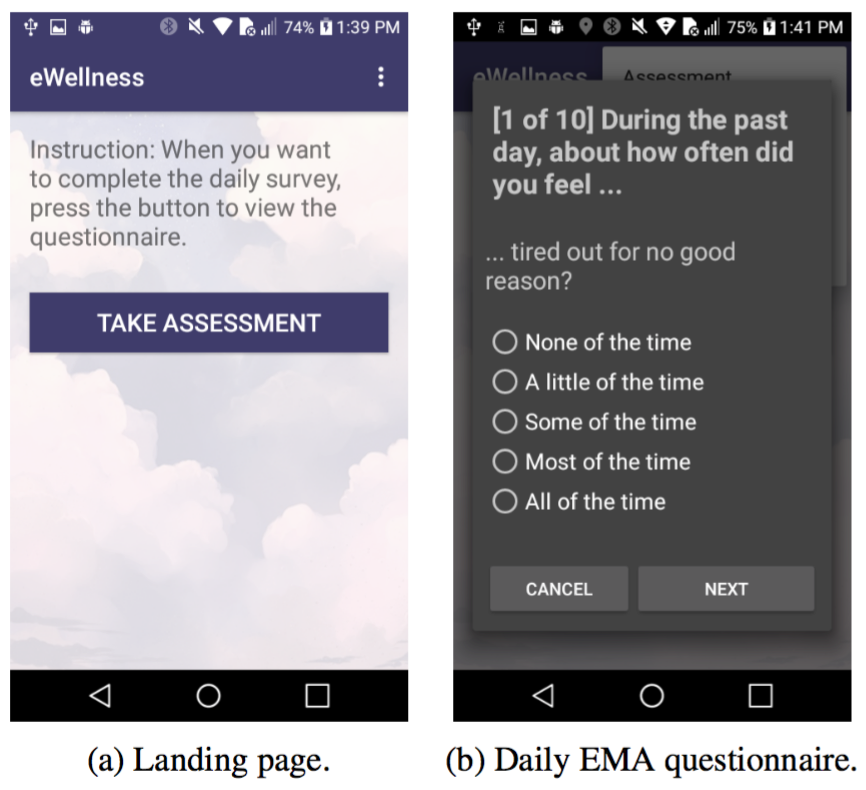}
  \caption{Screenshots of eWellness}
  \label{fig:app}
\end{figure}

An IRB-approved pilot study was conducted on a dozen individuals who are using the Android version $5.0$ and above from the university community, including students and staff. Study participants did not have a reported history of mental illness. Participants were asked to download and install the eWellness application (Figure~ \ref{fig:app}), and then run it on their phone for a month. Passive data was collected continuously by the application throughout the month. Participants were asked to answer EMA  daily through the eWellness app, but did not provide any other personal information, such as name, gender, age, during participation.

The Kessler Psychological Distress Scale (K10) \cite{andrews2001interpreting} is a validated measure of psychological distress during the past 30 days, which is used for clinical and epidemiological purposes. It has a notable success in measuring feelings of anxiety along with depression. For this pilot, the K10 was modified to assess criteria over the previous 24 hour period. The modified K10 prompted the users as daily EMA to measure their feelings of anxiety and depression. The K10 has ten items, which are scored from five through to one (all of the time, most of the time, some of the time, a little of the time, and none of the time). The minimum possible score of K10 is $10$, and the maximum possible score is $50$. K10 results are categorized into four levels of psychological distress: low distress, moderate distress, high distress, and very high distress (Table~\ref{tab:k10_results}). These results were leveraged as a label for the classification of supervised learning.

\begin{table}
\centering
  \caption{Categorization of K10 Scores \cite{australianstatistics}}
  \label{tab:k10_results}
  \begin{tabular}{ccc}
    \toprule
    K10 Score & Level & Samples (N=146) \\
    \midrule
    10-15 & Low distress & 91 \\
    16-21 & Moderate distress & 29 \\
    22-29 & High distress & 21 \\
    30-50 & Very high distress & 5 \\
  \bottomrule
\end{tabular}
\end{table}

\section{Results}
Only 10 participants answered at least seven days of EMAs and provided successful passive sensing data throughout the month. Our analysis focused on a fully supervised learning approach, and only labeled samples were included. For this pilot study, we used 146 daily samples to identify daily anxiety and depression levels. The Z-Score normalization was applied to the features to reach normalized values from different participants. 

% \begin{figure}
% \centerline{\includegraphics[width=0.45\textwidth]{figures/mental_tsne.png}}
%     \caption{tSNE mapping of classes.}
%     \label{fig:mental_tsme}
% \end{figure}

% T-distributed Stochastic Neighbor Embedding (t-SNE) \cite{maaten2008visualizing} provides additional insights as to the nature of the collected data. t-SNE represents each high-dimensional object to a two-dimensional point in such a way that nearby points model similar objects and distant points with high probability model dissimilar objects \cite{maaten2008visualizing}. In Figure \ref{fig:mental_tsme}, the vast majority of data points are red, which is low distress (level 0). While some of these level 0 data points are scattered, a distinct cluster of level 0 is perhaps segmented by gender type or mobile device usage type. 

We selected 25 features that have a relatively higher correlation with the raw K10 score. For the 4-class classification, we used 5-fold Cross-Validation (CV) with four models: K-Nearest Neighbors (KNN), Extra-Trees (ET), Support Vector Machine (SVM), and Multilayer Perceptron (MLP). The class weight was automatically applied to the models inversely proportional to the class frequencies to train the imbalanced dataset. The highest classification accuracy achieved was around 76\% with the extra-trees model. We also applied the under-sampling technique to improve the performance of an imbalanced dataset. Samples from the low distress class were removed randomly to make uniformly distributed class labels. Samples from the very high distress class were also ignored. A confusion-matrix demonstrates that the average score of classifying three classes is $0.65$.

\begin{figure}
\centerline{\includegraphics[width=0.30\textwidth]{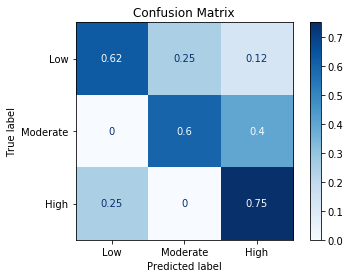}}
    \caption{3-class (Low, Moderate, and High distress) Classification Confusion Matrix.}
    \label{fig:mental_tsme}
\end{figure}

% \begin{table}
%   \caption{Classification Accuracy of 5-fold Cross-validation}
%   \label{tab:cv_result}
%   \begin{tabular}{cc}
%     \toprule
%     Model&Accuracy (SD) \\
%     \midrule
%     KNN & 0.681 (0.085) \\
%     ET & 0.760 (0.097) \\
%     SVM & 0.673 (0.083) \\
%     MLP & 0.656 (0.088) \\
%   \bottomrule
% \end{tabular}
% \end{table}

\section{Discussion}
\subsection{Limitations of the Study}
While 10 subjects completing one-months worth of continuous data represents a critical validation of the technology and 
its potential utility, the dataset is too small to achieve statistically significant results. Additionally, this pilot was scoped to only include individuals without a clinical diagnosis of Anxiety or Depression. Consequently, there were insufficient cases of user-reported mental distress, particularly moderate or severe cases, to classify effectively. Additional studies are planned to enlarge our dataset and include a cohort of individuals with diagnosed mental health conditions.

There is also significant concern about the veracity of user self-reported labeling of mental health. The experimental design, prompting users to fill out a daily EMA in the application via push-notification, encouraged active participation in the study; however, there was no mechanism for confirming that the resulting inputs were an accurate reflection of a user's actual wellbeing. Users were likely motivated to respond quickly, and not necessarily accurately, which likely resulted in default answers of no reported anxiety or depression. Furthermore, there may have been a reluctance among users to accurately report out mental health issues given perceived embarrassment or stigma associated with poor mental health.  The authors recommend that future studies will have to address these concerns by better anticipating and correcting for challenges with accurate labeling of mental health.  

Given the significant heterogeneity across subjects in-terms of usage-patterns, it was assumed that primary-success would be achieved by classifying within users across time, rather than across users. The limitations of this initial dataset did not allow for adequate classifying by individual; however, the fact that classification success was achieved by bundling samples across all subjects is remarkable in its indication that cross-subject learning in this domain could be possible. Part of this result likely stems from normalization performed on the data to account for habitual differences in subject usage. Additional data collection is necessary to validate this finding.

\subsection{Usability}

Attempting to gauge the viability of the concept, participants in the pilot were asked to submit a voluntary anonymized post-study questionnaire regarding their perceptions about the application and its data collection practices. All participants responded. A significant majority described the application as somewhat (40\%) or mostly (40\%) useful. Likewise, all users endorsed feeling comfortable with the application, and only one user expressed reservations about the data being collected. 

All participants obtained detailed accounting of the data that was collected as part of their onboarding process to the study. No individual declined to participate after learning the precise nature of what was being tracked. This sampling suggests that, particularly among the young adults who are more accustomed to digitized lives, there is less concern about data collection through their mobile devices. Limiting the collection of PII could be sufficient to assuage most privacy concerns.

The primary issue users had with the application was its battery consumption resulting from heavy over-sampling of the sensors. Future iterations of the application will seek to optimize battery usage by minimizing the sampling frequency.

\section{Conclusion}

Remote health monitoring of mental health, when done so leveraging passive and unobtrusive data collection, could be a useful alternative for conducting real-time mental health surveillance. This paper presents eWellness, an experimental mobile application designed to track a full-suite of sensor and log data off a user's device continuously and passively. An initial pilot study tracking ten people over a month showed a nearly 76\% success rate at predicting daily anxiety levels based solely on the passively monitored features. Our current approach may prove useful at tracking longitudinal trends in an individual's mental health. Additional work is needed to refine both the technology and analytics.

\bibliographystyle{./bibliography/IEEEtran}
\bibliography{./bibliography/IEEEabrv,./bibliography/IEEEexample,./bibliography/refs.bib}

\begin{thebibliography}{10}
\providecommand{\url}[1]{#1}
\csname url@samestyle\endcsname
\providecommand{\newblock}{\relax}
\providecommand{\bibinfo}[2]{#2}
\providecommand{\BIBentrySTDinterwordspacing}{\spaceskip=0pt\relax}
\providecommand{\BIBentryALTinterwordstretchfactor}{4}
\providecommand{\BIBentryALTinterwordspacing}{\spaceskip=\fontdimen2\font plus
\BIBentryALTinterwordstretchfactor\fontdimen3\font minus
  \fontdimen4\font\relax}
\providecommand{\BIBforeignlanguage}[2]{{%
\expandafter\ifx\csname l@#1\endcsname\relax
\typeout{** WARNING: IEEEtran.bst: No hyphenation pattern has been}%
\typeout{** loaded for the language `#1'. Using the pattern for}%
\typeout{** the default language instead.}%
\else
\language=\csname l@#1\endcsname
\fi
#2}}
\providecommand{\BIBdecl}{\relax}
\BIBdecl

\bibitem{bitsko2018epidemiology}
R.~H. Bitsko, J.~R. Holbrook, R.~M. Ghandour, S.~J. Blumberg, S.~N. Visser,
  R.~Perou, and J.~T. Walkup, ``Epidemiology and impact of health care
  provider--diagnosed anxiety and depression among us children,'' \emph{Journal
  of developmental and behavioral pediatrics: JDBP}, vol.~39, no.~5, p. 395,
  2018.

\bibitem{cartwright2006anxiety}
S.~Cartwright-Hatton, ``Anxiety of childhood and adolescence: Challenges and
  opportunities,'' 2006.

\bibitem{merikangas2010lifetime}
K.~R. Merikangas, J.-p. He, M.~Burstein, S.~A. Swanson, S.~Avenevoli, L.~Cui,
  C.~Benjet, K.~Georgiades, and J.~Swendsen, ``Lifetime prevalence of mental
  disorders in us adolescents: results from the national comorbidity survey
  replication--adolescent supplement (ncs-a),'' \emph{Journal of the American
  Academy of Child \& Adolescent Psychiatry}, vol.~49, no.~10, pp. 980--989,
  2010.

\bibitem{topper2017prevention}
M.~Topper, P.~M. Emmelkamp, E.~Watkins, and T.~Ehring, ``Prevention of anxiety
  disorders and depression by targeting excessive worry and rumination in
  adolescents and young adults: a randomized controlled trial,''
  \emph{Behaviour research and therapy}, vol.~90, pp. 123--136, 2017.

\bibitem{hammerness2008characterizing}
P.~Hammerness, T.~Harpold, C.~Petty, C.~Menard, C.~Zar-Kessler, and
  J.~Biederman, ``Characterizing non-ocd anxiety disorders in psychiatrically
  referred children and adolescents,'' \emph{Journal of affective disorders},
  vol. 105, no. 1-3, pp. 213--219, 2008.

\bibitem{APA}
A.~P. Association, ``What are anxiety disorders?''
  \url{https://www.psychiatry.org/patients-families/anxiety-disorders/what-are-anxiety-disordersg},
  2019.

\bibitem{newby2014effectiveness}
J.~M. Newby, L.~Mewton, A.~D. Williams, and G.~Andrews, ``Effectiveness of
  transdiagnostic internet cognitive behavioural treatment for mixed anxiety
  and depression in primary care,'' \emph{Journal of Affective Disorders}, vol.
  165, pp. 45--52, 2014.

\bibitem{bardone1998adult}
A.~M. Bardone, T.~E. Moffitt, A.~Caspi, N.~Dickson, W.~R. Stanton, and P.~A.
  Silva, ``Adult physical health outcomes of adolescent girls with conduct
  disorder, depression, and anxiety,'' \emph{Journal of the American Academy of
  Child \& Adolescent Psychiatry}, vol.~37, no.~6, pp. 594--601, 1998.

\bibitem{woodward2001life}
L.~J. Woodward and D.~M. Fergusson, ``Life course outcomes of young people with
  anxiety disorders in adolescence,'' \emph{Journal of the American Academy of
  Child \& Adolescent Psychiatry}, vol.~40, no.~9, pp. 1086--1093, 2001.

\bibitem{wagner2016mental}
S.~Wagner, C.~Koehn, M.~White, H.~Harder, I.~Schultz, K.~Williams-Whitt,
  O.~Warje, C.~Dionne, M.~Koehoorn, R.~Pasca \emph{et~al.}, ``Mental health
  interventions in the workplace and work outcomes: a best-evidence synthesis
  of systematic reviews,'' \emph{Int J Occup Environ Med (The IJOEM)}, vol.~7,
  no. 1 January, pp. 607--1, 2016.

\bibitem{american2016american}
A.~C.~H. Association \emph{et~al.}, ``American college health
  association--national health assessment ii: Reference group executive summary
  spring 2016,'' 2016.

\bibitem{bruffaerts2018mental}
R.~Bruffaerts, P.~Mortier, G.~Kiekens, R.~P. Auerbach, P.~Cuijpers,
  K.~Demyttenaere, J.~G. Green, M.~K. Nock, and R.~C. Kessler, ``Mental health
  problems in college freshmen: Prevalence and academic functioning,''
  \emph{Journal of affective disorders}, vol. 225, pp. 97--103, 2018.

\bibitem{swan2012health}
M.~Swan, ``Health 2050: The realization of personalized medicine through
  crowdsourcing, the quantified self, and the participatory biocitizen,''
  \emph{Journal of personalized medicine}, vol.~2, no.~3, pp. 93--118, 2012.

\bibitem{ciabattoni2017real}
L.~Ciabattoni, F.~Ferracuti, S.~Longhi, L.~Pepa, L.~Romeo, and F.~Verdini,
  ``Real-time mental stress detection based on smartwatch,'' in \emph{2017 IEEE
  International Conference on Consumer Electronics (ICCE)}.\hskip 1em plus
  0.5em minus 0.4em\relax IEEE, 2017, pp. 110--111.

\bibitem{osma2014proposal}
J.~Osma, I.~Plaza, E.~Crespo, C.~Medrano, and R.~Serrano, ``Proposal of use of
  smartphones to evaluate and diagnose depression and anxiety symptoms during
  pregnancy and after birth,'' in \emph{IEEE-EMBS International Conference on
  Biomedical and Health Informatics (BHI)}.\hskip 1em plus 0.5em minus
  0.4em\relax IEEE, 2014, pp. 547--550.

\bibitem{ben2015next}
D.~Ben-Zeev, E.~A. Scherer, R.~Wang, H.~Xie, and A.~T. Campbell,
  ``Next-generation psychiatric assessment: Using smartphone sensors to monitor
  behavior and mental health.'' \emph{Psychiatric rehabilitation journal},
  vol.~38, no.~3, p. 218, 2015.

\bibitem{narziev2020stdd}
N.~Narziev, H.~Goh, K.~Toshnazarov, S.~A. Lee, K.-M. Chung, and Y.~Noh, ``Stdd:
  Short-term depression detection with passive sensing,'' \emph{Sensors},
  vol.~20, no.~5, p. 1396, 2020.

\bibitem{fukazawa2019predicting}
Y.~Fukazawa, T.~Ito, T.~Okimura, Y.~Yamashita, T.~Maeda, and J.~Ota,
  ``Predicting anxiety state using smartphone-based passive sensing,''
  \emph{Journal of biomedical informatics}, vol.~93, p. 103151, 2019.

\bibitem{andrews2001interpreting}
G.~Andrews and T.~Slade, ``Interpreting scores on the kessler psychological
  distress scale (k10),'' \emph{Australian and New Zealand journal of public
  health}, vol.~25, no.~6, pp. 494--497, 2001.

\bibitem{australianstatistics}
\BIBentryALTinterwordspacing
``Chapter - k10 scoring.'' [Online]. Available:
  \url{https://www.abs.gov.au/ausstats/abs@.nsf/lookup/4817.0.55.001Chapter92007-08}
\BIBentrySTDinterwordspacing

\end{thebibliography}

\end{document}